\documentclass[11pt]{article}
\usepackage{authblk,graphicx,caption2,subfigure}
\makeatletter
\newcommand{\singlespacing}{\let\CS=\@currsize\renewcommand{\baselinestretch}{1.0}\tiny\CS}
\newcommand{\doublespacing}{\let\CS=\@currsize\renewcommand{\baselinestretch}{1.25}\tiny\CS}

\begin{document}
\title{Transverse Mass Distribution Characteristics of $\pi^0$ Production
in $^{208}$Pb-induced Reactions and the Combinational Approach}

\author{Bhaskar De\thanks{e-mail: bhaskar$\_$r@isical.ac.in}  and S.
Bhattacharyya\thanks{e-mail: bsubrata@isical.ac.in
(Communicating Author).}}

\affil{Physics
and Applied Mathematics Unit(PAMU),\\
Indian Statistical Institute, Kolkata - 700108, India.}

\date{\today}

\maketitle

\begin{abstract}
The nature of invariant cross-sections and multiplicities in some
$^{208}Pb$-induced reactions and some important ratio-behaviours of the
invariant multiplicities for various centralities of the collision
will here be dealt with in the light of a combinational approach which
has been built up in the recent past by the present authors. Next, the
results would be compared with the outcome of some of the simulation-based
standard models for multiple production in nuclear collisions at high energies.
Finally, the implications of all this would be discussed.\\


\end{abstract}

\newpage

\section{Introduction}

The reasons for interests in the specialized studies on
$Pb$-induced reactions are modestly well-known and are, by now,
somewhat commonplace for which we are not going to elaborate on
them. In the recent past the WA98 Collaboration\cite{Aggarwal1}
presented a detailed study of neutral pion transverse mass spectra
in the range $0.5 GeV/c^2\leq m_T - m_0 \leq 4.0 GeV/c^2$ and $2.3
\leq y \leq 3.0$ for collisions of $PbPb$ and $PbNb$ at $158A$ GeV
for different centralities. Besides, Aggarwal et al (WA98
Collaboration)\cite{Aggarwal1} also pointed out how the data on
the ratios of invariant multiplicities in nucleus-nucleus
reactions to those in proton-proton reactions diverge from values
predicted by nearly all the important standard versions of
multiple production of particles in nucleus-nucleus collisions.
Against this perspective, we would try to understand here the
reported features of neutral pion production in two
lead($A=208)$-induced reactions at $158A$ GeV and also try to
explain the problematical ratios in the light of a combinational
approach\cite{De1}. The present study is conducted obviously with
the clear motivation of testing this newly projected approach for
understanding the characteristics of particle production in heavy
ion collisions.
\par
We would concentrate here, in the main, on the following few
aspects: (i) the nature of pionic inclusive cross-section, (ii)
the property of average transverse momentum, (iii) the qualitative
character of the multiplicity-behaviour and of their ratios, (iv)
the nature of centrality-dependence of the ratios in the various
$p_T$-intervals. Our results on these observables and the
available data would also be compared with some of the most
prominent models for multiparticle production in heavy ion
collisions. The paper would be organized as follows: The section 2
offers the outline of the model and the sketch of the physical
ideas which prompted us to proceed in the stated direction. In
section 3 we present the calculational results and their graphical
descriptions. The last section is, as usual, for our final
comments on the totality of the work presented here.

\section{The Model: A Sketch}
Following the suggestion of Faessler\cite{Faessler1} and the work
of Peitzmann\cite{Peitzmann1} and also of Schmidt and
Schukraft\cite{Schmidt1}, we propose here a generalized empirical
relationship between the inclusive cross-section for pion
production in nucleon(N)-nucleon(N) collision and that for
nucleus(A)-nucleus(B) collision as given below:

\begin{equation}
\displaystyle{E\frac{d^3  \sigma}{dp^3} ~ (AB \rightarrow \pi X) ~
\sim ~ (AB)^{\phi(y, ~ p_T)} ~ E\frac{d^3  \sigma}{dp^3} ~ (PP
\rightarrow \pi X) ~,} \label{eqn1}
\end{equation}

where $\phi(y, ~ p_T)$ could be expressed in the factorization
form, $\phi(y, ~ p_T) = f(y) ~ g(p_t)$; and the product, $AB$ on
the right hand side of the above equation is the product of mass
numbers of the two nuclei participating in the collisions at high
energies, of which one will be the projectile and the other one
the target.
\par
While investigating any specific nature of dependence of the two
variables ($y$ and $p_T$), either one of these is assumed to
remain constant. In other words, more particularly, if and when
the $p_t$-dependence is studied by experimental groups, the
rapidity factor is treated to be constant and the vice-versa. So,
the formula for studying the nature of $p_T$-spectra turns into

\begin{equation}
\displaystyle{E\frac{d^3 \sigma}{dp^3} ~ (AB \rightarrow \pi X) ~
\sim ~ (AB)^{g(p_T)} ~ E\frac{d^3  \sigma}{dp^3} ~ (PP \rightarrow
\pi X) ~,} \label{eqn2}
\end{equation}

The main bulk of work, thus, converges to the making of an appropriate
choice of form for $g(p_T)$. And the necessary choices are to be made
on the basis of certain premises and physical considerations which do
not violate the canons of high energy particle interactions.
\par
The expression for inclusive cross-section of pions in
proton-proton scattering at high energies in Eqn.(2) could be
chosen in the form suggested first by Hagedorn\cite{Hagedorn1}:
\begin{equation}
\displaystyle{ E\frac{d^3  \sigma}{dp^3} ~ (PP \rightarrow \pi X)
~ = ~ C_1 ~ ( ~ 1 ~ + ~ \frac{p_T}{p_0})^{-n} ~ ,} \label{eqn3}
\end{equation}

where $C_1$ is the normalization constant, and $p_o$, $n$ are
interaction-dependent chosen phenomenological parameters for which
the values are to be obtained by the method of fitting.
\par
The final working formula for the nucleus-nucleus collisions is
now being proposed here in the form given below:

\begin{equation}
\begin{array}{lcl}
E\frac{d^3  \sigma}{dp^3} ~ (AB \rightarrow \pi X) ~ & \propto & ~
(AB)^{(\epsilon ~ + ~ \alpha  p_T  ~ - ~ \beta p_T^2)} ~
E\frac{d^3 \sigma}{dp^3} ~ (PP \rightarrow \pi X) \\
& \propto & ~ (AB)^{(\epsilon ~ + ~ \alpha  p_T  ~ - ~ \beta
p_T^2)} ~ (1 ~ + ~ \frac{p_T}{p_0})^{-n} ~ ,  \label{eqn4}
\end{array}
\end{equation}

with $ g(p_T) ~ = ~ (\epsilon ~ + ~ \alpha  p_T  ~ - ~ \beta
p_T^2)$, where this suggestion of quadratic parametrization for
$g(p_T)$ is exclusively made by us and is called hereafter
De-Bhattacharyya parametrization(DBP). In the above expression
$\epsilon$, $\alpha$ and $\beta$ are constants for a specific pair
of projectile and target.
\par
Earlier experimental works\cite{Aggarwal1,Albrecht1,Antreasyan1}
showed that $g(p_T)$ is less than unity in the $p_T$-domain,
$p_T<1.5$ GeV/c. Besides, it was also observed that the parameter
$\epsilon$, which gives the value of $g(p_T)$ at $p_T=0$, is also
less than one and this value differs from collision to collision.
The other two parameters $\alpha$ and $\beta$ essentially
determine the nature of curvature of $g(p_T)$. However, in the
present context, precise determination of $\epsilon$ is not
possible for the following understated reasons:
\par
(i) To make our point let us recast the expression for (4) in the
form given below:
\begin{equation}
\displaystyle{E\frac{d^3\sigma}{dp^3}(AB \rightarrow \pi X) ~
\approx ~ C_2 ~ (AB)^\epsilon ~ (AB)^{(\alpha p_T - \beta p_T^2)}
~ ( ~ 1 ~ + \frac{p_T}{p_0} ~ )^{-n}}\label{eqn5}
\end{equation}
where $C_2$ is the normalization term which has a dependence
either on the rapidity or on the rapidity density of the pion and
which also absorbs the previous constant term,$C_1$ as well.
\par
Quite obviously, we have adopted here the method of fitting. Now,
in Eqn.(5) one finds that there are two constant terms $C_2$ and
$\epsilon$ which are neither the coefficients nor the exponent
terms of any function of the variable, $p_T$. And as $\epsilon$ is
a constant for a specific collision at a specific energy, the
product of the two terms $C_2$ and $(A.B)^\epsilon$ appears as
just a new constant. And, it will just not be possible to obtain
fit-values simultaneously for two constants of the above types by
the method of fitting.
\par
(ii) From Eqn.(2) the nature of $g(p_T)$ can easily be determined
by calculating the ratio of the logarithm of the ratios of
nuclear-to-$PP$ collision and the logarithm of the product $AB$.
Thus, one can measure $\epsilon$ from the intercept of $g(p_T)$
along y-axis as soon as one gets the values of
$E\frac{d^3\sigma}{dp^3}$ for both $AB$ collision and $PP$
collision at the same c.m. energy. But, there is a problem that it
will not be possible to get readily  the data on inclusive spectra
for $PP$ collisions at all c.m.energy values.
\par
In order to sidetrack these difficulties and also to build up an
escape-route, we have concentrated here almost wholly to the values of
$\alpha$ and $\beta$ for various collision systems and the
resultant effects of $C_2$ and $\epsilon$ have been absorbed into
a single constant term $C_3$. Hence, the final expression becomes
\begin{equation}
\displaystyle{E\frac{d^3\sigma}{dp^3}(AB \rightarrow \pi X) ~
\approx ~ C_3 ~ (AB)^{(\alpha p_T - \beta p_T^2)} ~ ( ~ 1 ~ +
\frac{p_T}{p_0} ~ )^{-n}}\label{eqn6}
\end{equation}
with $C_3 = C_2 (AB)^\epsilon$.
\par
The exponent factor term $\alpha p_T - \beta p_T^2$ obviously
represents here $[g(p_T)-\epsilon]$ instead of $g(p_T)$ alone. The
expression(6) given above is the physical embodiment of what we
have termed to be the grand combination of models(GCM) that has
been utilized here. The results of $PP$ scattering are obtained in
the above on the basis of eqn.(3) provided by Hagedorn's
model(HM);  and the route for converting the results of $NN$ to
$NA$ or $AB$ collisions is built up by the Peitzmann's
approach(PA) represented by expression(2). The further input is
the De-Bhattacharyya parametrization for the nature of the
exponent. Thus, the GCM is the combination of HM, PA
and the DBP, all of which are used here.
\par
And the choice of this form of parametrization for the power of
the exponent in eqn.(4) is not altogether a coincidence. In
dealing with the EMC effect in the lepton-nucleus collisions, one
of the authors here(SB),\cite{SB1} made use of a polynomial form
of $A$-dependence with the variable $x_F$(Feynman Scaling
variable). This gives us a clue to make a similar choice for both
$g(p_T)$ and $f(y)$ variable(s) in each case separately. In the
recent times, De-Bhattacharyya parametrization is being
extensively applied to interpret the measured data on the various
aspects\cite{De2} of the particle-nucleus and nucleus-nucleus
interactions at high energies. In the recent past Hwa et.
al.\cite{Hwa1} also made use of this sort of relationship in a
somewhat different context. The underlying physics implications of
this parametrization stem mainly from the expression(4) which
could be identified as a clear mechanism for switch-over of the
results obtained for nucleon-nucleon($PP$) collision to those for
nucleus-nucleus interactions at high energies in a direct and
straightforward manner. The polynomial exponent of the product
term on $AB$ takes care of the totality of the nuclear effects.
\par
For the sake of clarity and confirmation, let us further emphasize
a point here very categorically. It is to be noted that this
model(GCM) containing all the Eqns.(4), (5) and (6) was described
in some detail earlier and was made use of in analyzing extensive
sets of data in the previous publicatioins\cite{De1,De3} by the
same authors. And in verifying the validity of this model further,
the purpose here is to apply the same model to some other
problematical aspects of data which we would dwell upon in the
subsequent sections.
\par
Indeed, quite obviously, there are two phenomenological parameters
in $g(p_T)$ which need to be physically explained and/or
identified. In compliance with this condition we offer the
following physical explanations for the occurrence of all these
factors. The particle-nucleus or nucleus-nucleus collisions at
high energies almost instantly gives rise, {\it{ex hypothesi}}, to
an expanding blob or fireball with rising temperature. In real and
concrete terms this stage indicates the growing participation of
the already-expanded nuclear blob. As temperature increases at
this stage, the emission of highly energetic secondaries(which are
mostly peripheral nucleons or baryons) with increasing transverse
momentum is perfectly possible. The coefficient $\alpha$ addresses
this particularity of the natural event; and this is manifested in
the enhancement of the nuclear contribution with the rise of the
transverse momentum. Thereafter, there is a turnabout in the state
of reality. After the initial fractions of seconds, the
earlier-excited nuclear matter starts to cool down and there is a
clear natural contraction at this stage as the system suffers a
gradual fall in temperature. Finally, this leads to what one might
call `freeze-out' stage, which results in extensive hadronization,
especially in production of hadrons with very low transverse
momentum. In other words, the production of large-$p_T$ particles
at this stage is lowered to a considerable extent. This fact is
represented by the damping or attenuation term for the production
of high-$p_T$ particles. The factor $\beta$ with negative values
takes care of this state of physical reality. Thus the function
denoted by $g(p_T)$ symbolizes the totality of the features of the
expansion-contraction dynamical scenario in the after-collision
stage. This interpretation is, at present, is only suggestive.
However, let us make some further clarifications.
\par
The physical foundation that has here been attempted to be built
up is inspired by thermodynamic pictures, whereas the quantitative
calculations are based on a sort of pQCD-motivated power-law
formula represented by eqn.(4). This seems to be somewhat
paradoxical, because it would be hard to justify the hypothesis of
local thermal equilibrium in multihadron systems produced by high
energy collisions in terms of successive collision of the
QCD-partons(like quarks and gluons) excited or created in the
course of the overall process. Except exclusively for central
heavy ion collisions, a typical parton can only undergo very few
interactions before the final-state hadrons `freeze out', i.e.
escape as free particles or resonances. The fact is the hadronic
system, before the freeze-out starts, expands a great deal -- both
longitudinally and transversally -- while these very few
interactions take place\cite{Hove1}. But the number of parton
interactions is just one of the several other relevant factors for
the formation of local equilibrium. Of equal importance is the
parton distribution produced early in the collision process. This
early distribution is supposed to be a superposition of collective
flow and highly randomized internal motions in each space cell
which helps the system to achieve a situation close to the
equilibrium leading to the appropriate values of collective
variables including concerned and/or almost concerned quantities.
The parameter $\alpha$ in expression(4) is a measure of the ratio
of the net binary collision number to the total permissible number
among the constituent partons in the pre-freeze out expanding
stage identified to be a sort of explosive
`detonation'\cite{Hove1} stage. This is approximated by a
superposition of collective flow and thermalized internal motion,
which is a function of hadronic temperature manifested in the
behaviour of the average transverse momentum. The post freeze-out
hadron production scenario is taken care of by the soft
interaction which is proportional\cite{Aggarwal1,Li1} to the
number of participant nucleons, $N_{part}$, according to almost
any variety of wounded nucleon model. The factor $\beta$, we
conjecture, offers a sort of the ratio of the actual participating
nucleons to the total number of maximum allowable(participating)
nucleons. In fact, this sort of physical explanations seems to
have been acceded to by some of the physics community through
their approval of some of our previous works\cite{De2,De3}.

\section{The Calculations and the Results}

\begin{figure}
    \centering
        \includegraphics[width=8cm]{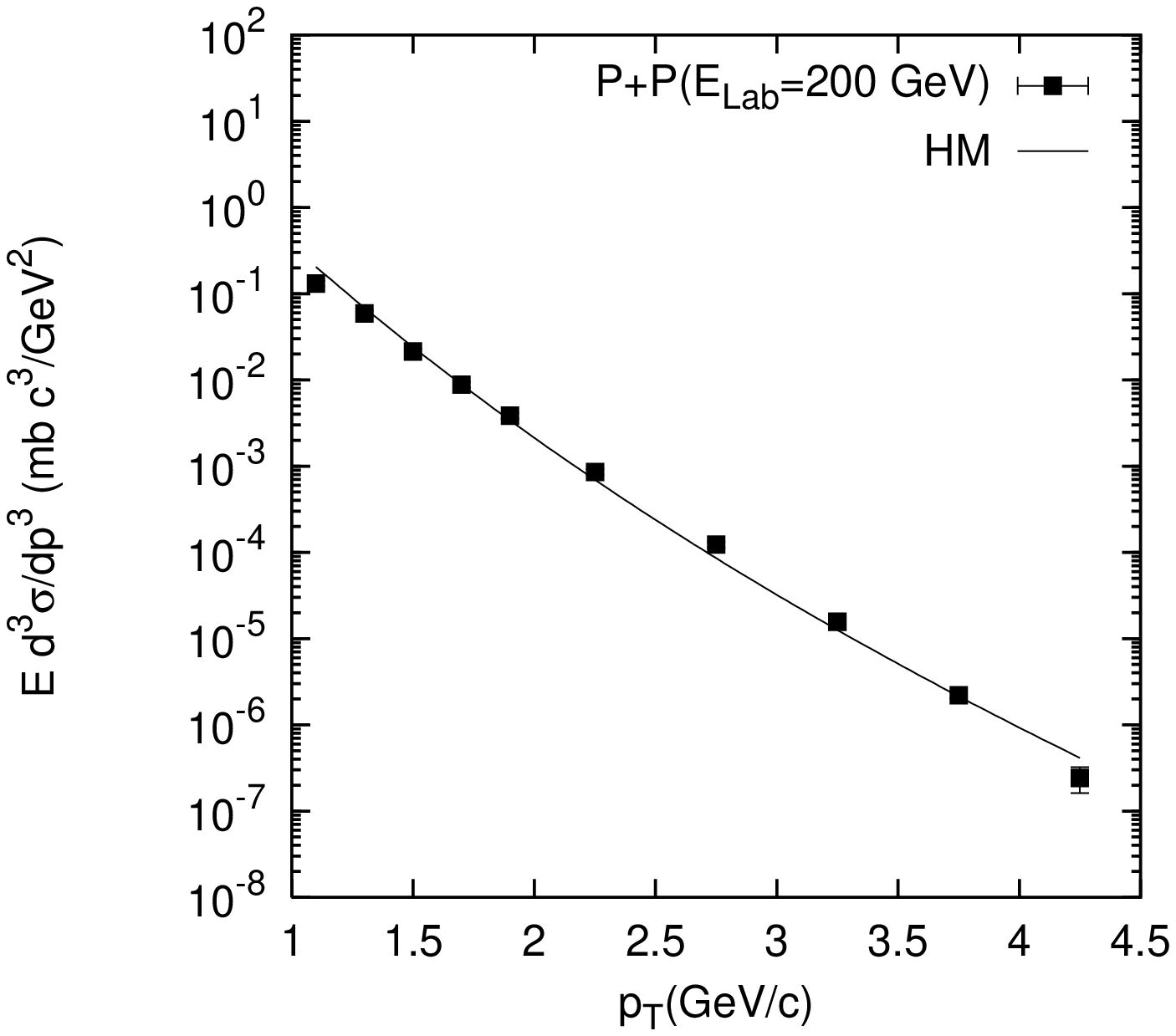}
    \caption{Plot of $E\frac{d^3\sigma}{dp^3}$ as a function of
transverse momentum, $p_T$ for production of secondary neutral
pions in $PP$ collision at $E_{Lab}=200$ GeV. The experimental
data are taken from \cite{Donaldson1}. The solid curve represents
the fit obtained on the basis of eqn.(3).}
    \label{pp200}
        \includegraphics[width=8cm]{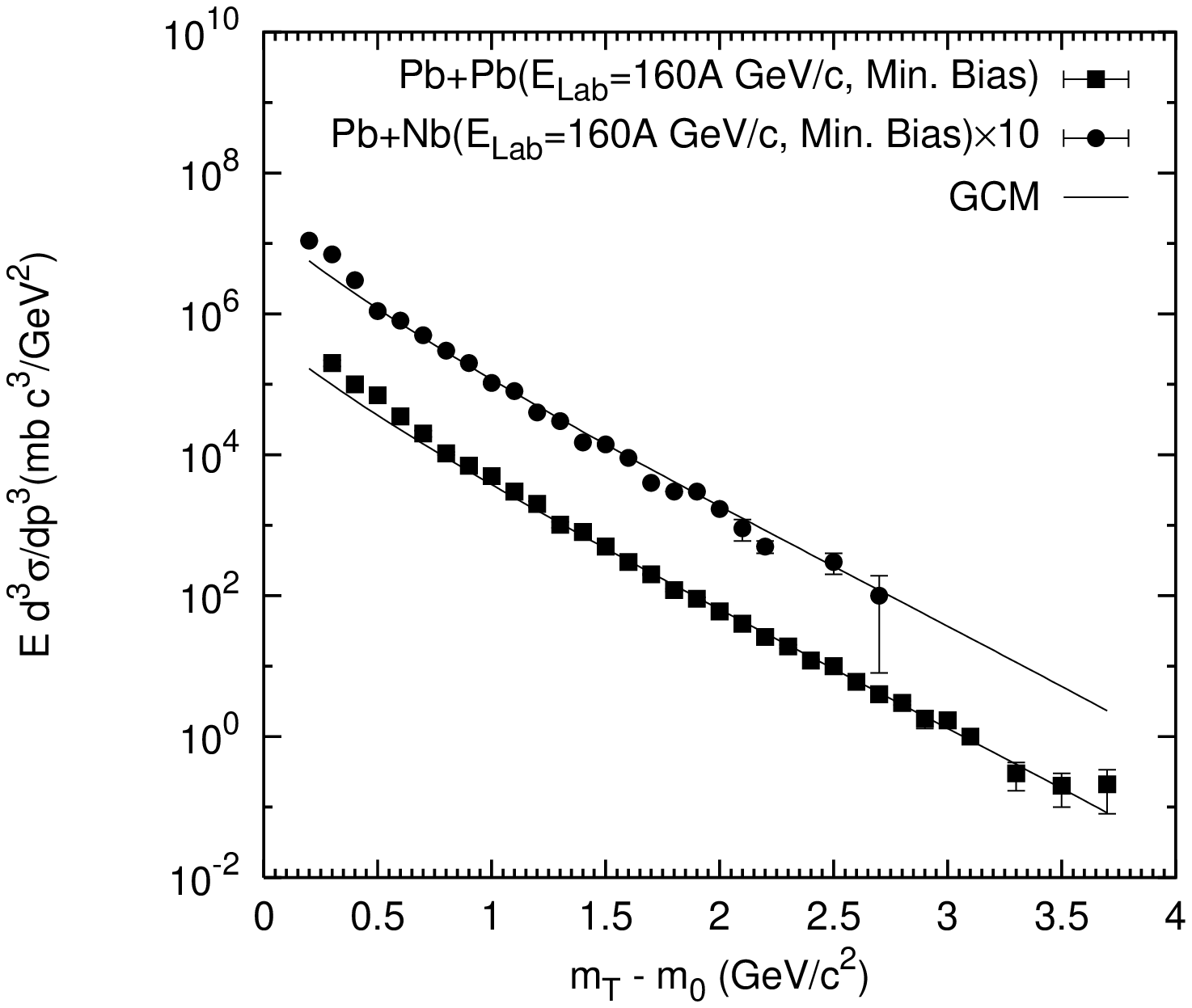}
    \caption{The inclusive spectra of secondary neutral pions
produced in two $Pb-$induced reactions at $E_{Lab}=160A$
GeV(minimum bias). All the experimental data are taken from
\cite{Aggarwal1}. The GCM-based results(eqn.(6)) are shown by the
solid curves.}
    \label{pbn160mb}
\end{figure}

The graphical plots presented in the diagrams in Fig.~\ref{pp200}
to Fig.~\ref{allclasses} describe the measured data on pion
production modestly well in the region of the moderate values of
the transverse momentum. The Fig.1 presents the fit for data on
$PP$ collisions. As described in the previous section the
parameter values to be inducted in calculations of $PP$ cross
sections are to be obtained from Table-1. These are based on
Hagedorn's model. Besides, in order to arrive at the theoretical
values of invariant pion production cross section in $PbPb$ and
$PbNb$ collisions at the measured energy with the help of this
combinational approach, one has got to use the fit values of
$\alpha$ and $\beta$ as provided in Table-2 and Table-3 for
minimum bias events and for the various centralities of the
nuclear collisions respectively. The average number of binary
nucleon-nucleon collisions, $<N_{coll}>$ and that of participant
nucleons, $<N_{part}>$ at different centralities are to be
obtained from Table-4 which is an adaptation of a specific set of
simulation results\cite{Aggarwal1}. The statement on the nature of
agreement with regard to the invariant multiplicity shown in
Fig.~\ref{allclasses} for pion production in $PbPb$ collisions at
$158A$ GeV in the most peripheral and most central collisions, and
also in the minimum bias events remains valid. These results
obtained by the GCM are compared in Fig.~\ref{modelratio} with
some of the very popular models in the field, while the
comparisons with the extracted data\cite{Aggarwal1} as well lay in
the background in all cases. Compatibility with data over a wide
range of the $p_T$-values is modestly satisfactory. The graphs on
the nature of the average transverse momentum vs. the number of
participating nucleons which is a measure of the centrality of the
collisions, depict faithfully (Fig.~\ref{averagept}) the expected
behaviours. The average transverse momentum values here are
normally defined by:

\begin{equation}
\displaystyle{<p_T> ~ = ~ \frac{\int_{p_T^{min}}^{p_T^{max}} {p_T
~ \frac{dN}{dp_T} ~
dp_T}}{\int_{p_T^{min}}^{p_T^{max}}{\frac{dN}{dp_T} ~ dp_T}}}\label{eqn7}
\end{equation}
where $p_T^{min}$ and $p_T^{max}$ are the lower and upper limits
of the transverse momentum, $p_T$ over which the integration has
been performed. The values of average transverse momenta, $<p_T>$
have been calculated in three $m_T-m_0$ intervals; and they are
$0.5 ~ GeV/c^2\leq m_T-m_0\leq 2.0 ~ GeV/c^2$, $1.0 ~ GeV/c^2\leq m_T-m_0\leq 2.0
~ GeV/c^2$ and $2.0 ~ GeV/c^2\leq m_T-m_0\leq 3.0 ~ GeV/c^2$. Here $m_T$
and $m_0$ are the transverse mass and the rest mass of secondary
pions respectively; and they are related with $p_T$ by the
equation $m_T^2=p_T^2+m_0^2$.
\par
However, for a specific nucleus-nucleus collision, the nature of
dependence of any observable on the center of mass(c.m.) energy
per basic colliding pair of nucleons, i.e.$\sqrt{s_{\rm{NN}}}$, is
to be obtained from the same for what it is in
nucleon-nucleon($PP$) reactions at high energies.
\par
The ratio-behaviours of the $p_T$-spectra and the nature of
another observable as would be defined later are studied here in
the rest of the diagrams. The ratio-values are of two types. In
Fig.~\ref{classratio} the ratio-values of the inclusive
$p_T$-spectra for various degree of centrality of collisions to
the same for minimum bias events have been studied.  The other set
of ratios are for the multiplicities for various
centrality-degrees of collisions. The diagrams shown in
Fig.~\ref{ratio1}-\ref{ratio4} demonstrate the plots of the ratios
of invariant multiplicity distributions of neutral pions
normalized to the number of binary collisions versus the
transverse momentum.  In Fig.~\ref{ratio1} the ratio of the
invariant cross section of neutral pions produced in least central
$PbPb$ collision($N_{Coll} \approx 10$) normalized by the number
of binary collisions to that of neutral pions produced in $PP$
collisions has been plotted as a function of transverse momentum;
and the ratio-values, in this particular case, are given by

\begin{equation}
\displaystyle{Ratio(\frac{Pb+Pb}{P+P}) ~ \equiv ~
\frac{\frac{E\frac{d^3N}{dp^3}}{N_{coll}}|_{N_{coll}=10}}
{E\frac{d^3N}{dp^3}|_{PP}}}
\label{eqn8}
\end{equation}

In the figure(Fig.~\ref{ratio1}) the extracted data shown by the
filled squares were obtained from Aggarwal et al\cite{Aggarwal1}
who assumed some parametrization for $PP$ reaction. In case of
GCM, in order to obtain the values of $E\frac{d^3N}{dp^3}|_{PP}$,
we have first utilized the parameter values given in Table-1 for
the $E\frac{d^3\sigma}{dp^3}|_{PP}$ and then normalized them by
the total inelastic cross section $\sigma_{in}= 43.6 \pm 4.0$ mb
\cite{Adamus1}, according to the conventional rule of conversion
arising out of the definitions.
\par
The diagrams in Fig.~\ref{ratio2}-\ref{ratio4} depict the ratios
of invariant cross sections at different centralities with a
normalization of the corresponding number of binary collisions.
The ratios in these cases can be written in the form

\begin{equation}
\displaystyle{Ratio(\frac{N_{coll}=X}{N_{coll}=Y}) ~ \equiv ~
\frac{\frac{E\frac{d^3N}{dp^3}}{N_{coll}}|_{N_{coll}=X}}
{\frac{E\frac{d^3N}{dp^3}}{N_{coll}}|_{N_{coll}=Y}}}
\label{eqn9}
\end{equation}

where $X$ and $Y$ denote two different number of binary collisions
at two different centralities.
\par
The observable plotted in all the diagrams of
Fig.~\ref{rbin1}-\ref{rbin4} and denoted by $R_{bin}$, is defined
by the relation ,

\begin{equation}
\displaystyle{R_{bin} ~ \equiv ~ E \frac{d^3N}{dp^3} / N_{coll}}
\end{equation}
The extracted data\cite{Aggarwal1} on the observable related to
neutral pion yield per binary collisions are for different
centralities in the separate $p_T$-intervals.
\par
Both the sets of figures(Fig.~\ref{ratio1}-\ref{ratio4} and
Fig.~\ref{rbin1}-\ref{rbin4}) provide roughly satisfactory
description of the data. So, the present work could be presented
as a successful continuation of one of our previous
work\cite{De1}. Thus all this lends substantial degree of credence
to the combinational approach built up and completed by us. And in
comparison with even the most accepted version of the
simulation-based standard models, i.e. the HIJING model, the
present approach works quite well.

\section{Concluding Remarks}

\begin{figure}
    \centering
        \includegraphics[width=9cm]{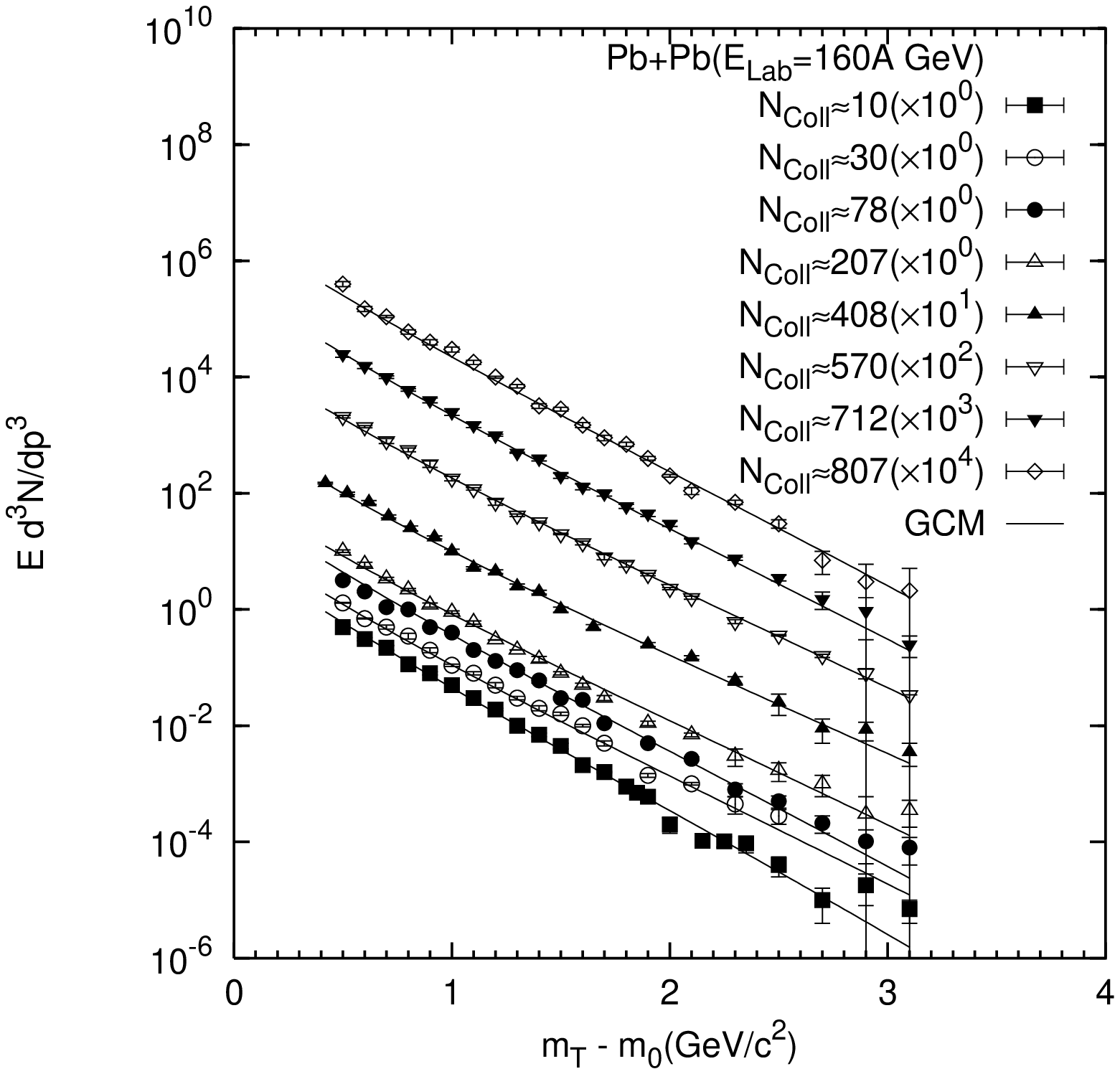}
    \caption{Invariant multiplicities of $\pi^0$ produced in $Pb+Pb$
collisions of different centralities at $E_{Lab}=160A$ GeV as a
function of $m_T-m_0$. The experimental data-points for various
centrality bins are taken from \cite{Aggarwal1}. The solid curves
provide the GCM-based fits(eqn.(6)).}
    \label{allclasses}
        \includegraphics[width=9cm]{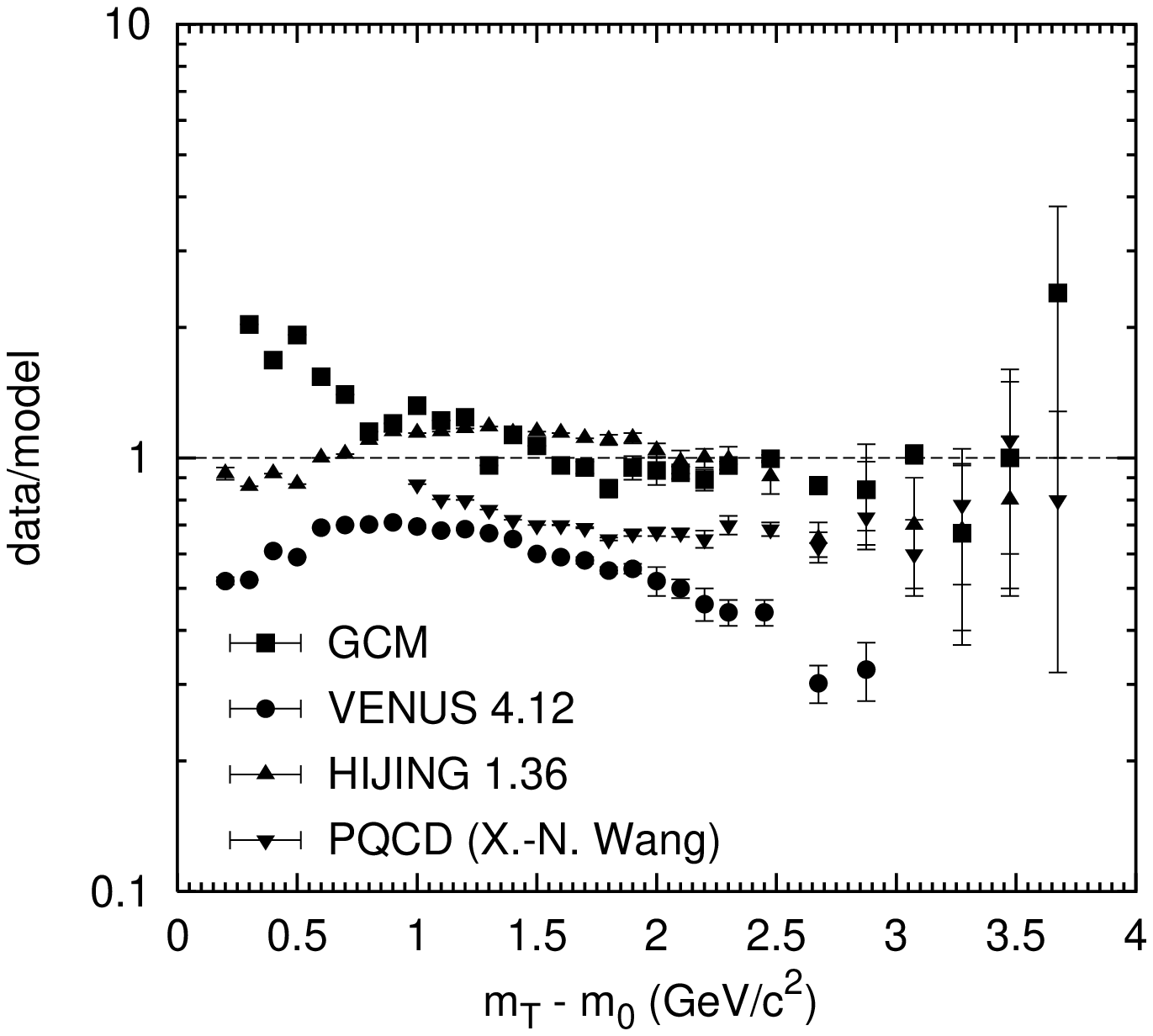}
    \caption{A comparison of the performance of different models via the
ratio of the minimum bias data for $Pb+Pb$ collision to
model-based results. The solid squares provide the GCM-based
results. Other model-based results are taken from
\cite{Aggarwal1}.}
    \label{modelratio}
\end{figure}

\begin{figure}
 \centering
        \includegraphics[width=8cm]{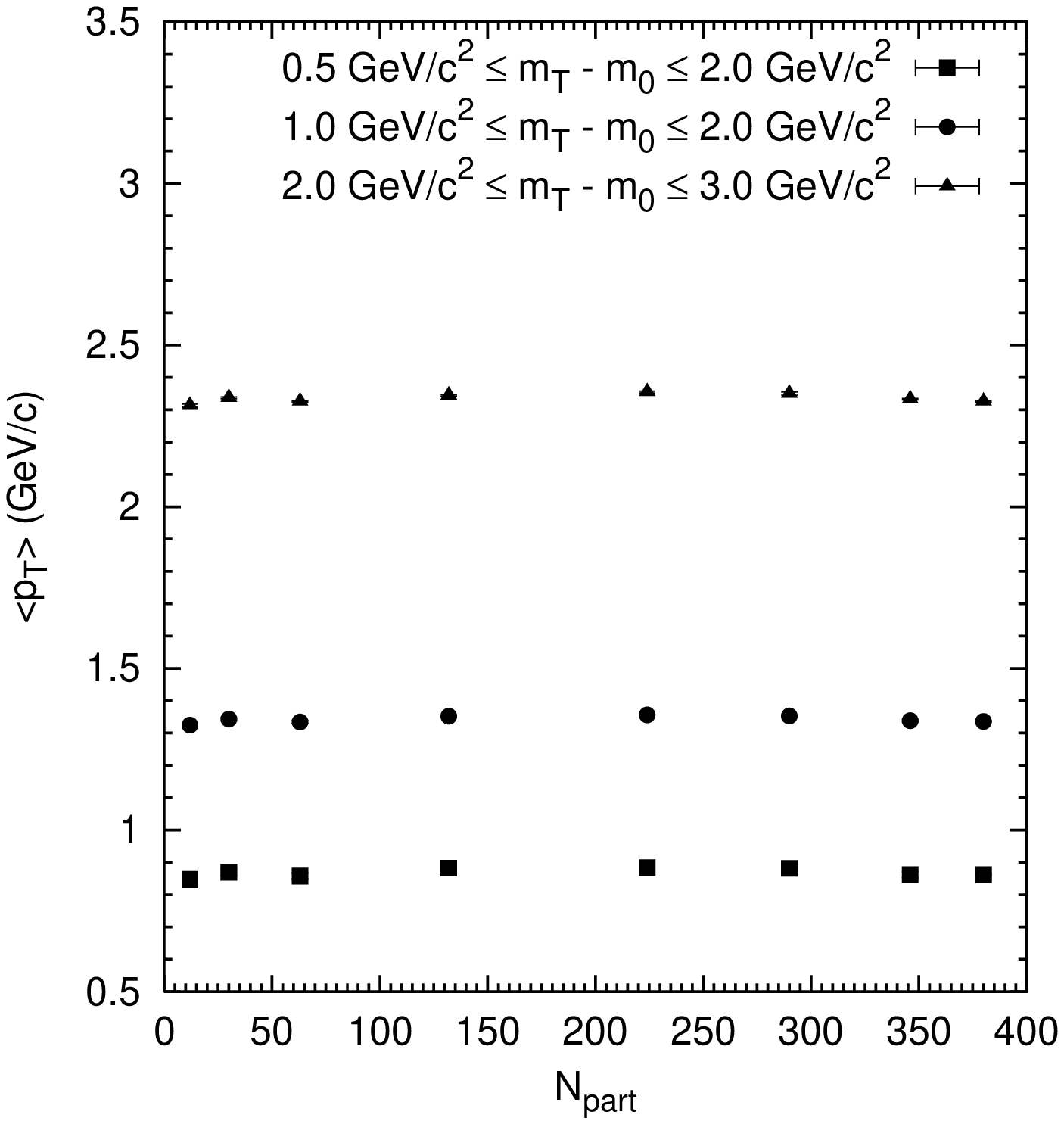}
    \caption{Plot of $<p_T>$, obtained on the basis of the present
model(GCM), as a function of the no. of participant nucleons,
$N_{part}$.}
    \label{averagept}
        \includegraphics[width=8cm]{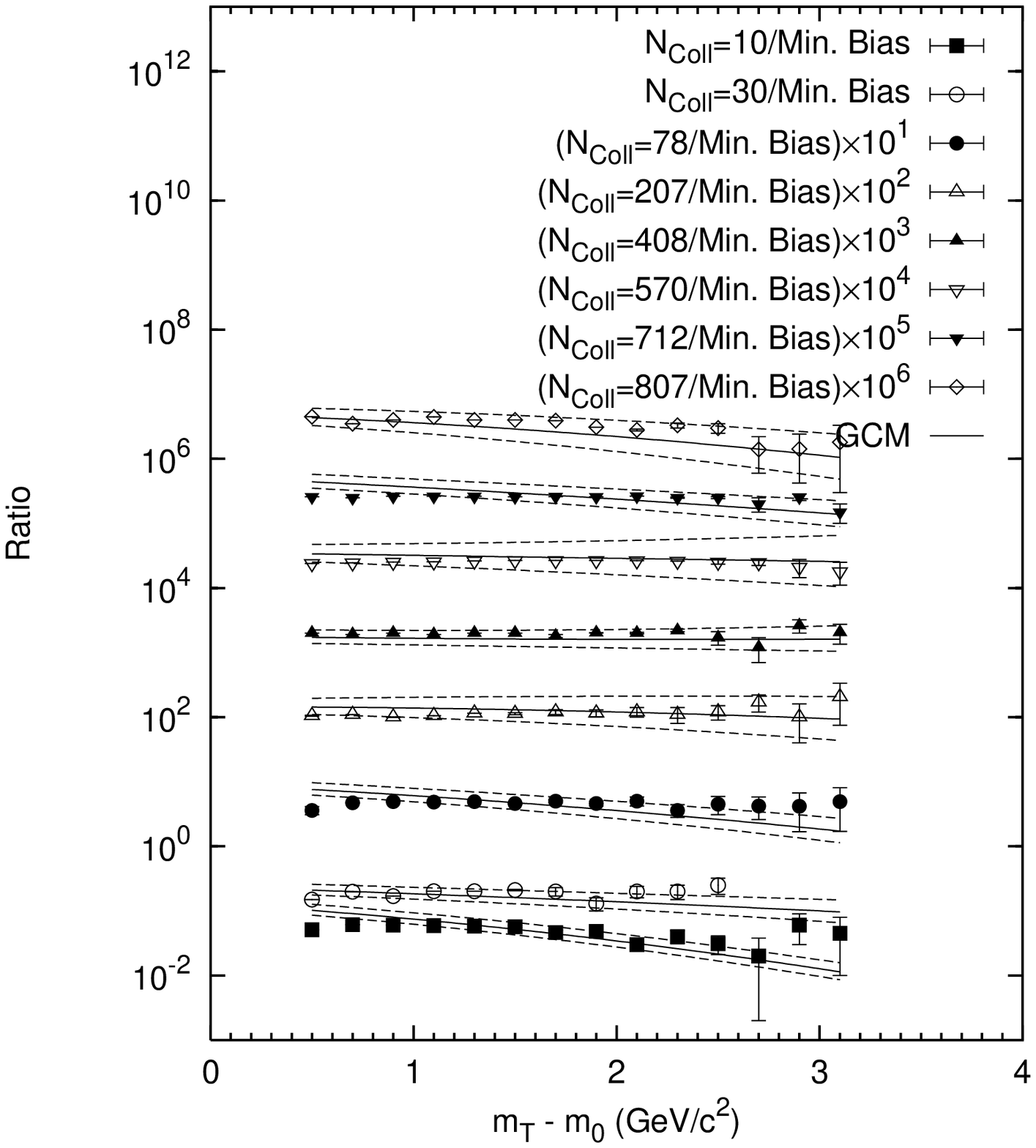}
    \caption{Plot of ratios of invariant multiplicities of neutral pions for
$Pb+Pb$ collisions of different centralities to minimum bias
events. The data-type points represent the experimentally measured
values while the solid curvilinear lines show the GCM-based
results. The dashed curves are for uncertainties arising out of
the error-ranges of all the parameter values.}
    \label{classratio}
    \end{figure}

\begin{figure}
    \subfigure[]{
    \begin{minipage}{.5\textwidth}
    \centering
        \includegraphics[width=8cm]{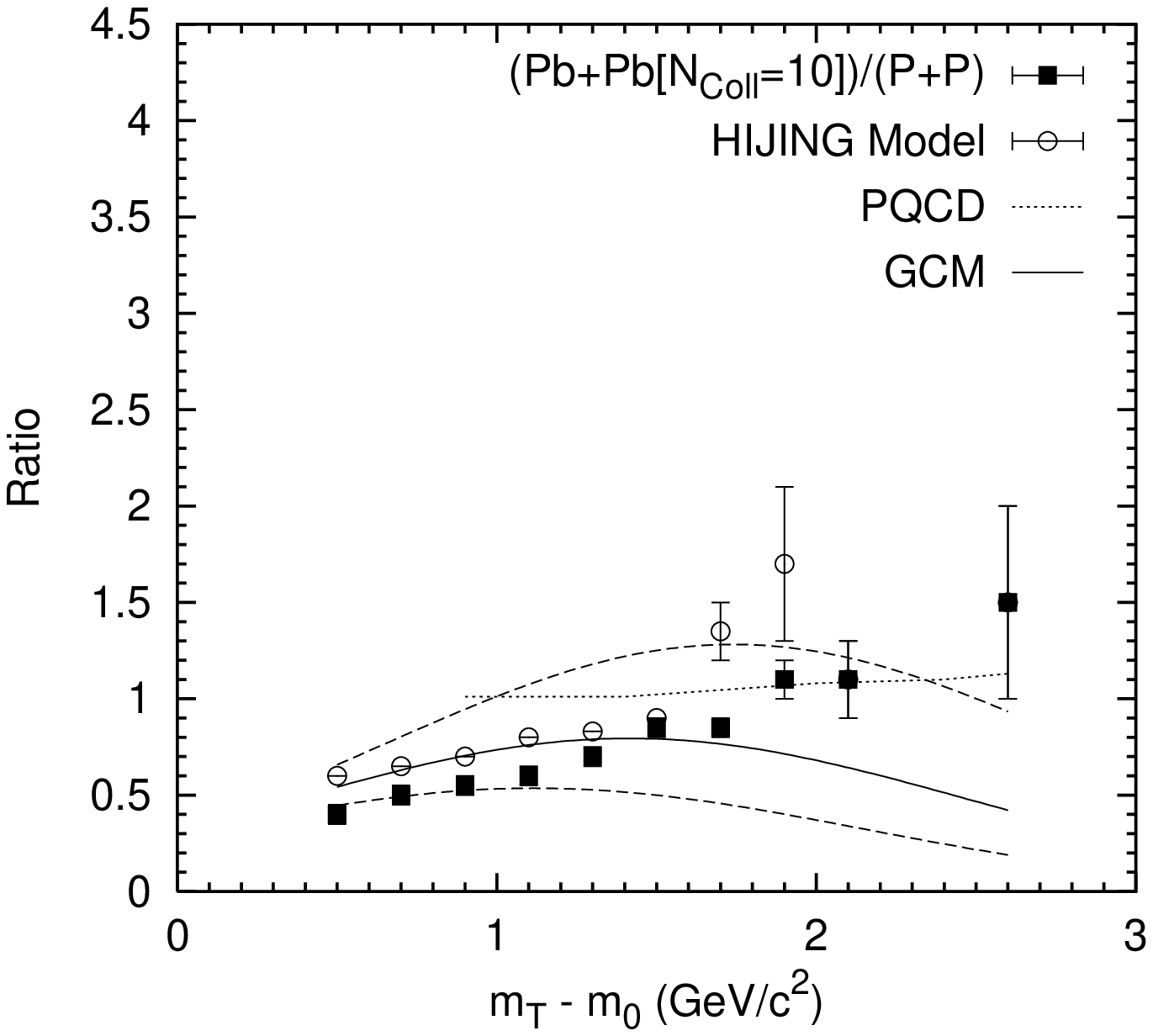}
        \label{ratio1}
    \end{minipage}}%
  \subfigure[]{
    \begin{minipage}{.5\textwidth}
    \centering
        \includegraphics[width=8cm]{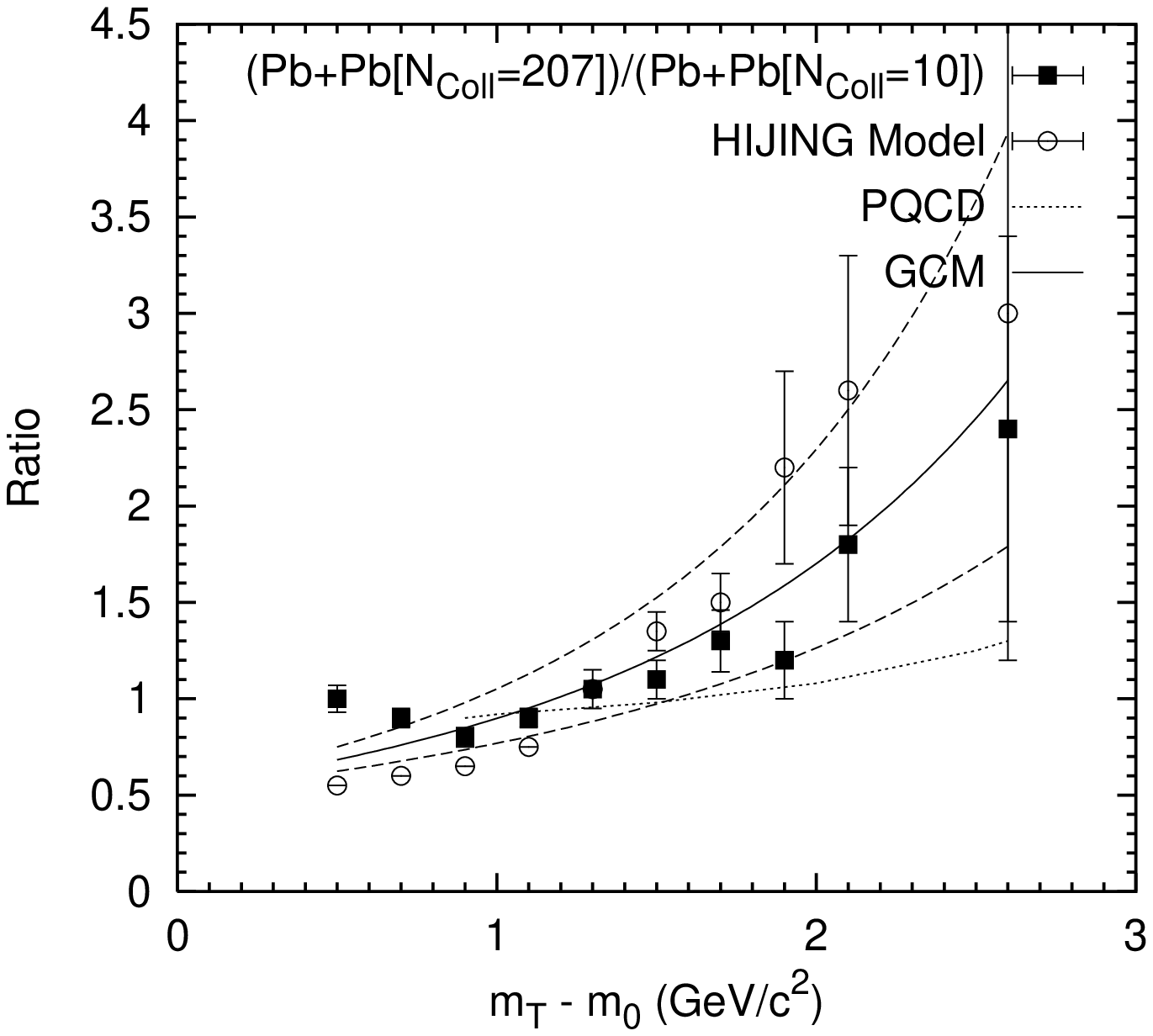}
        \label{ratio2}
    \end{minipage}}%
  \vspace{.1cm}
  \subfigure[]{
  \begin{minipage}{.5\textwidth}
    \centering
        \includegraphics[width=8cm]{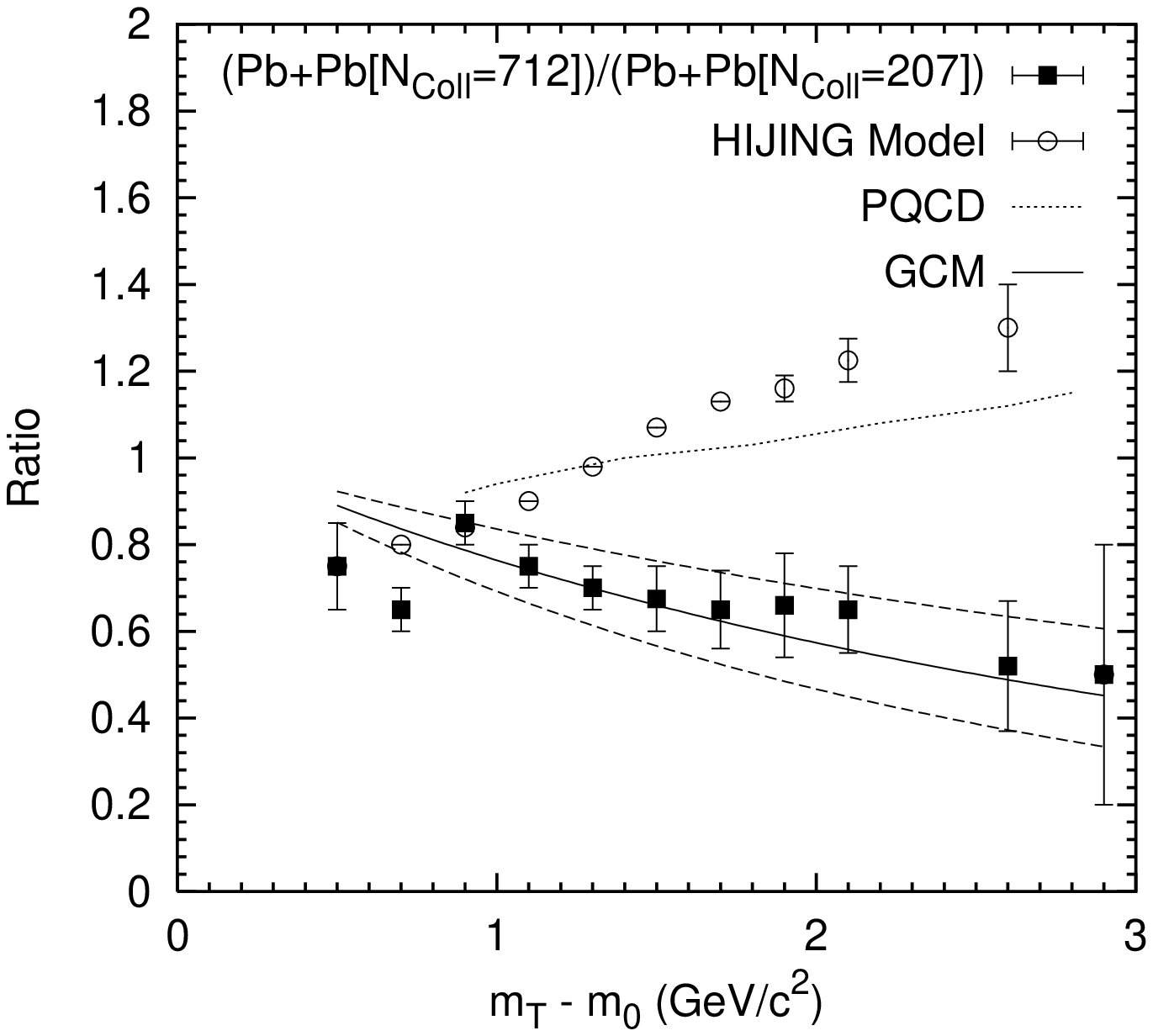}
        \label{ratio3}
    \end{minipage}}%
  \subfigure[]{
  \begin{minipage}{.5\textwidth}
    \centering
        \includegraphics[width=8cm]{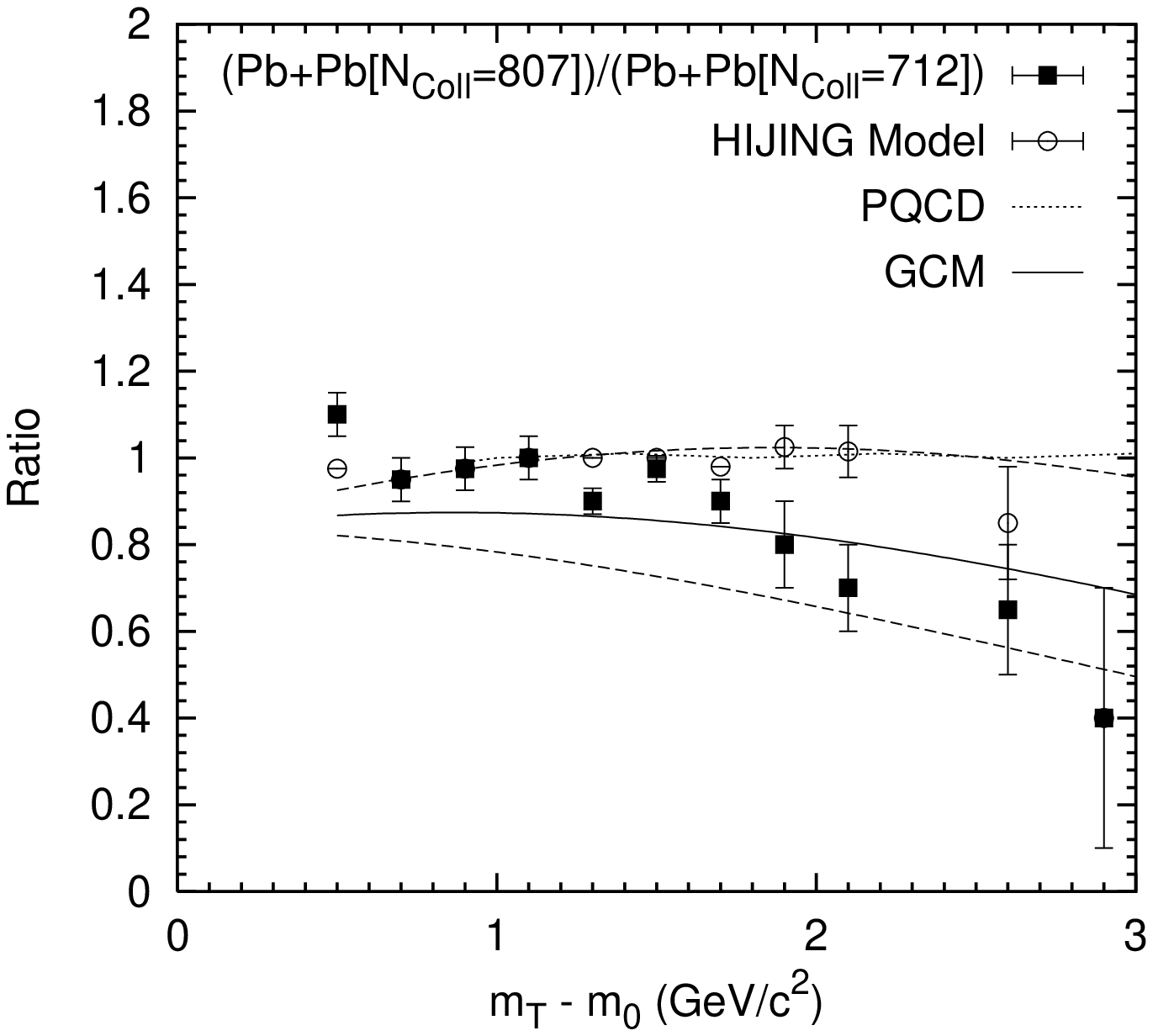}
        \label{ratio4}
    \end{minipage}}%
  \label{ratio}
 \caption{Ratios of $E\frac{d^3N}{dp^3}$ of neutral pions for different
centralities normalized to the number of binary collisions. The
GCM-based results are depicted by the solid curves. The dashed
curves represent the uncertainties arising due to the errors in
$C_3$, $\alpha$ and $\beta$. Other model-based results have been
obtained from \cite{Aggarwal1}.}
\end{figure}

\begin{figure}
    \subfigure[]{
    \begin{minipage}{.5\textwidth}
    \centering
        \includegraphics[width=8cm]{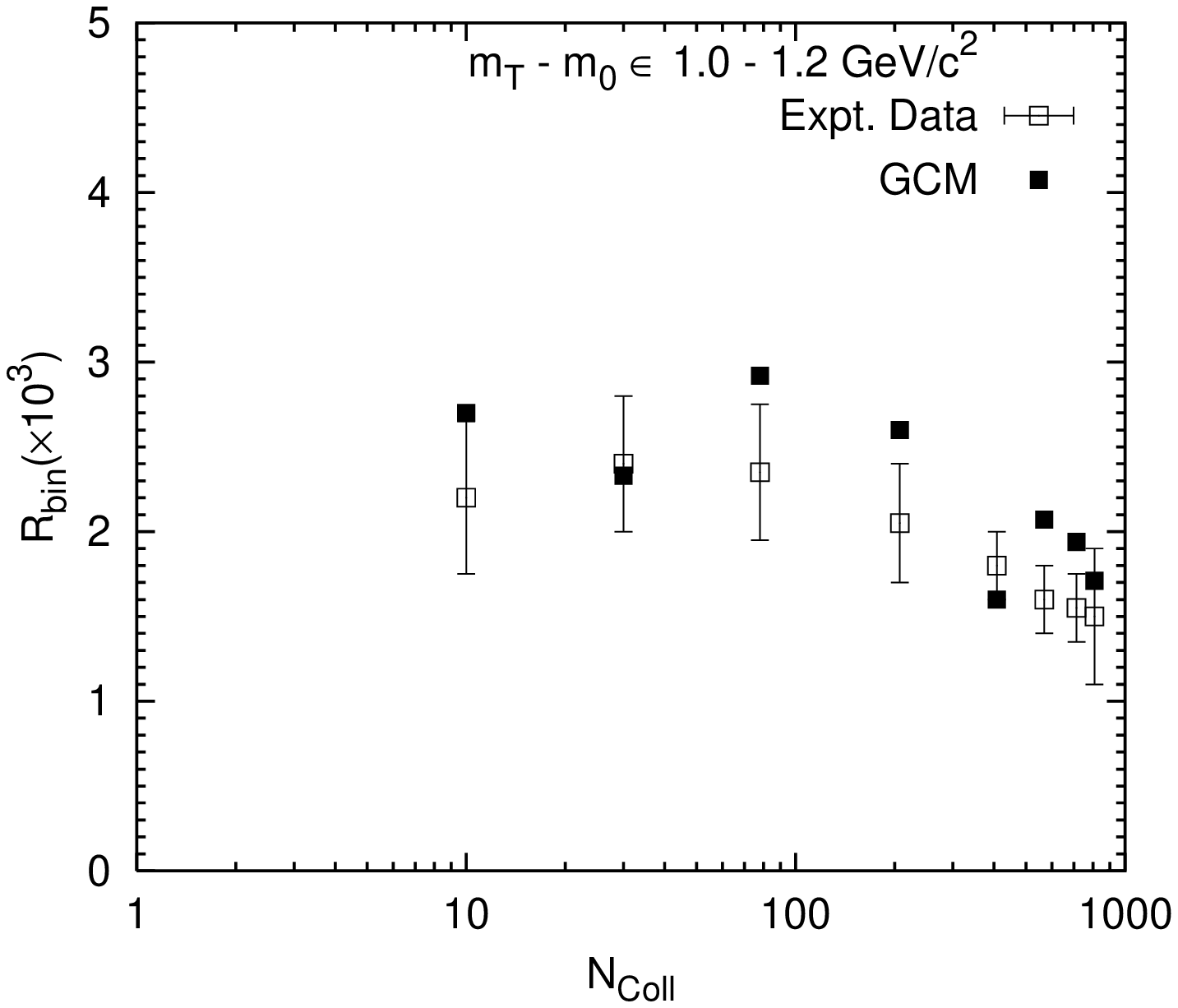}
        \label{rbin1}
    \end{minipage}}%
  \subfigure[]{
    \begin{minipage}{.5\textwidth}
    \centering
        \includegraphics[width=8cm]{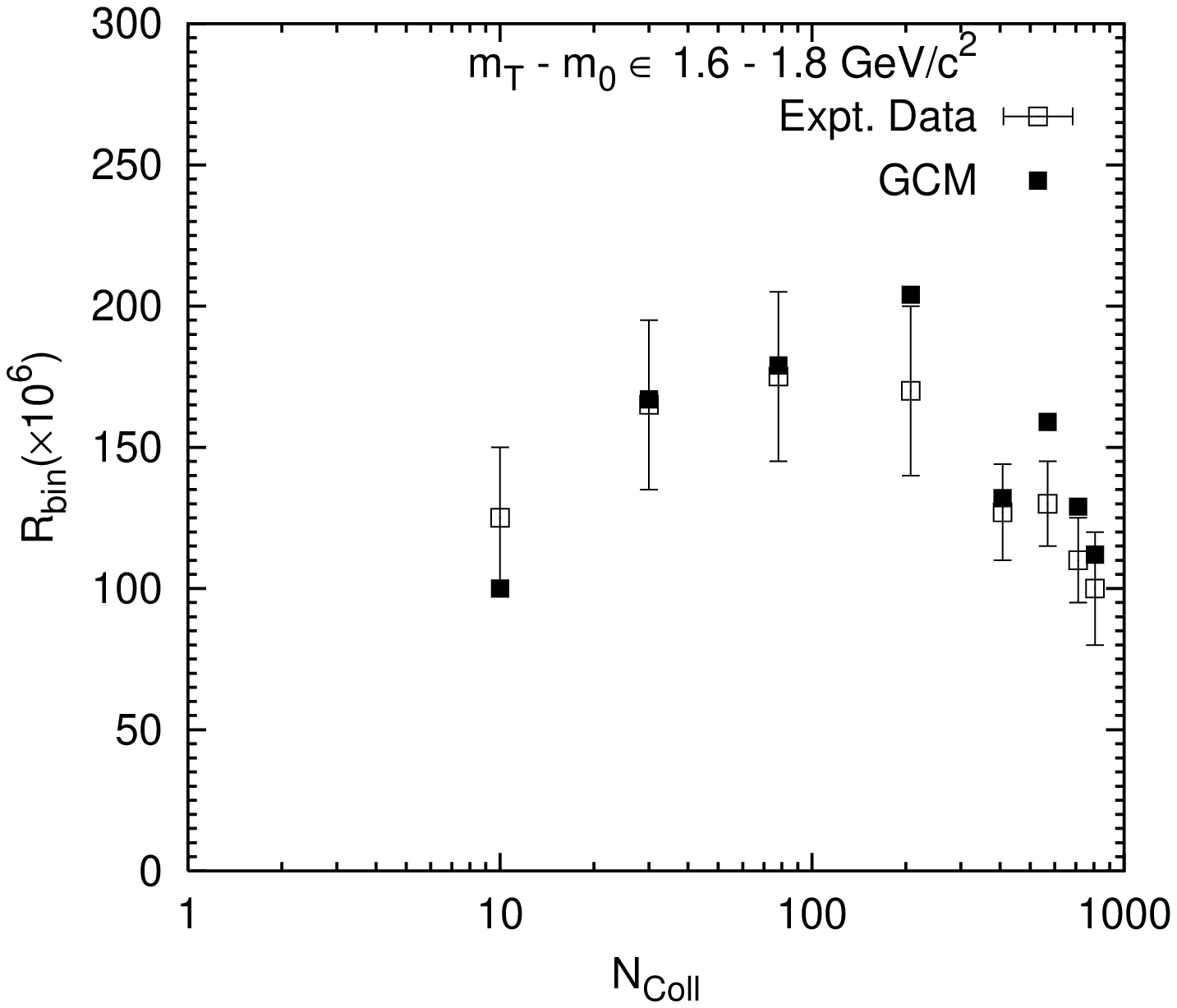}
        \label{rbin2}
    \end{minipage}}%
  \vspace{.1cm}
  \subfigure[]{
  \begin{minipage}{.5\textwidth}
    \centering
        \includegraphics[width=8cm]{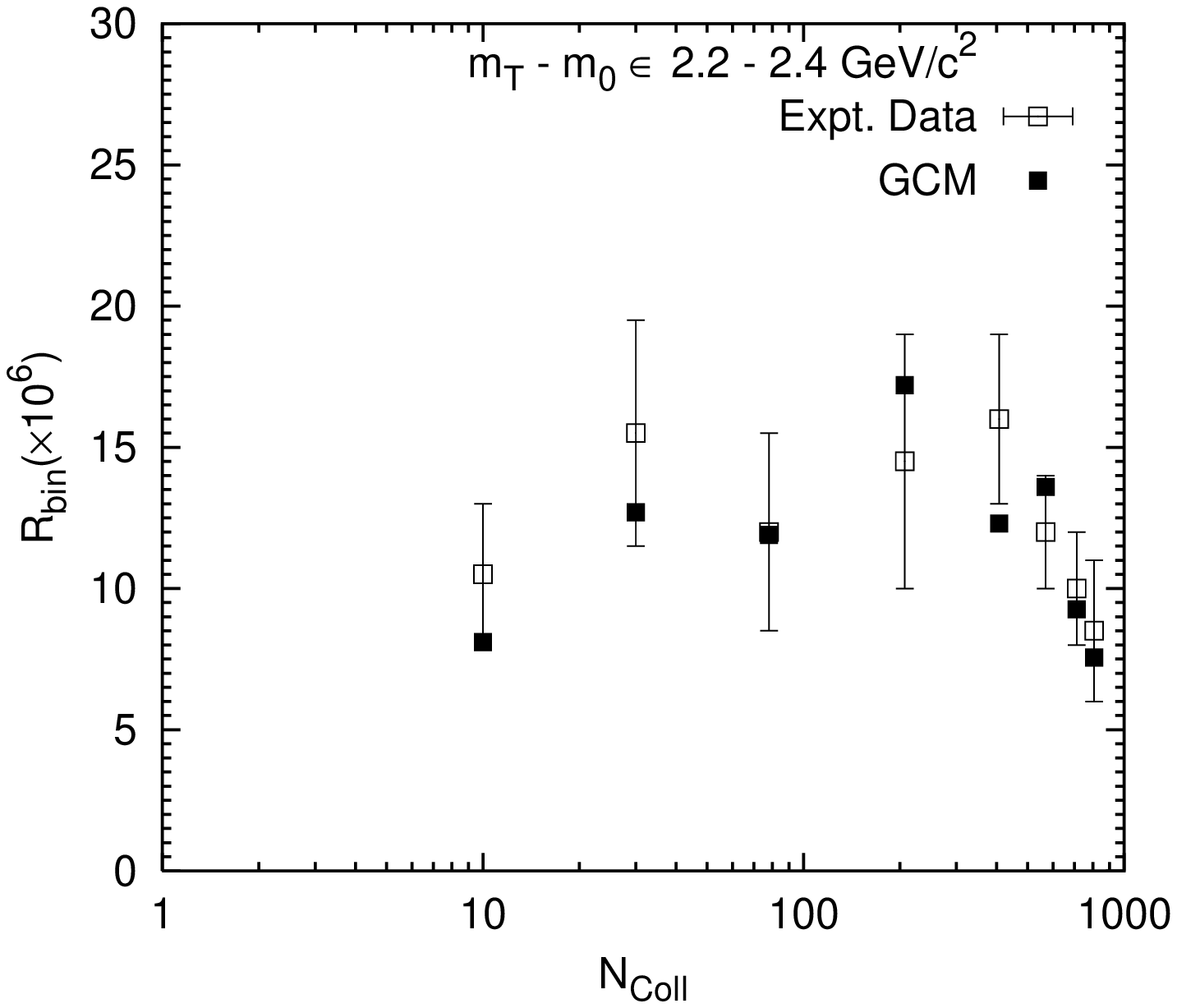}
        \label{rbin3}
    \end{minipage}}%
  \subfigure[]{
  \begin{minipage}{.5\textwidth}
    \centering
        \includegraphics[width=8cm]{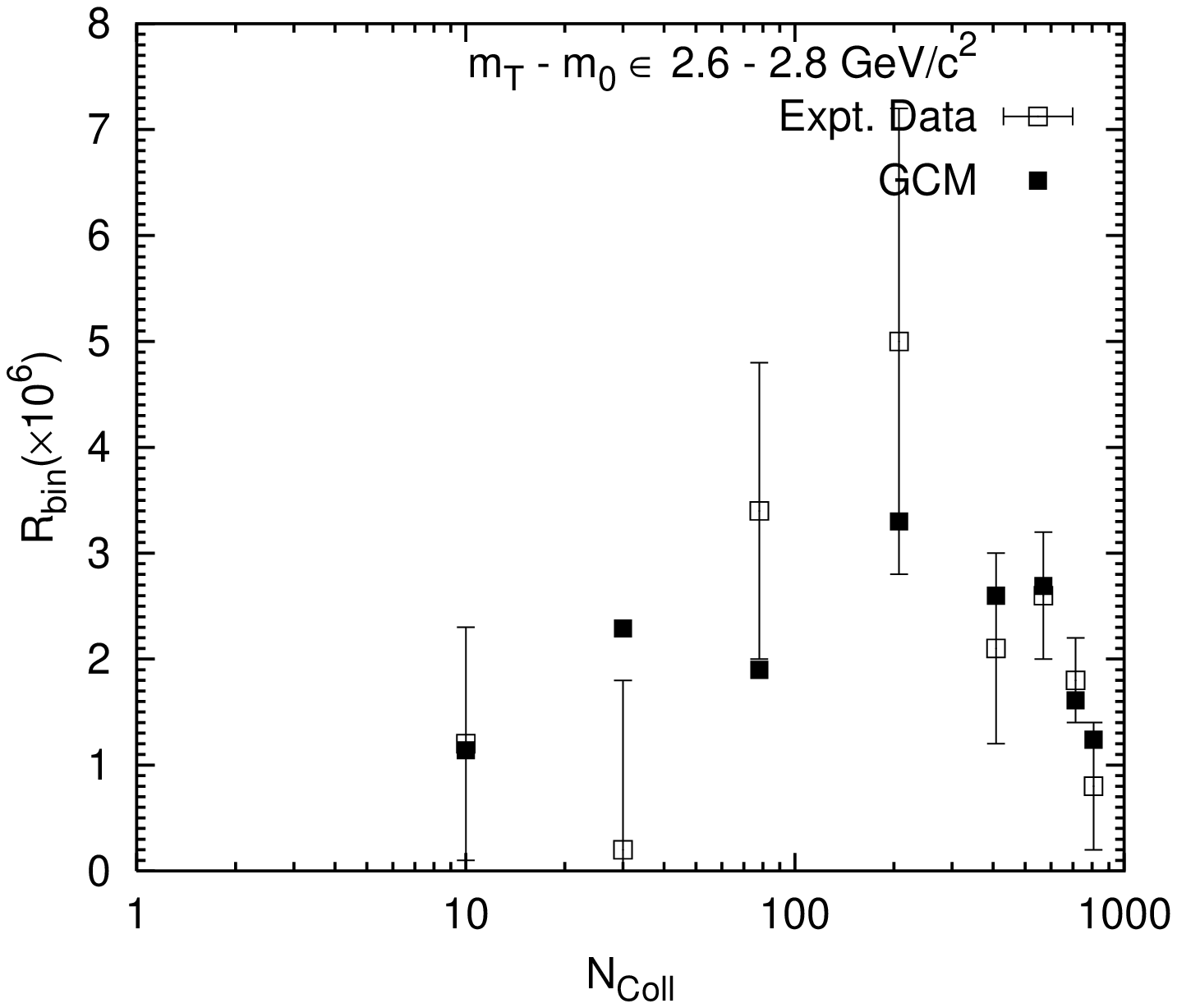}
        \label{rbin4}
    \end{minipage}}%
  \label{rbin}
 \caption{Invariant multiplicities of neutral pions normalized to the number
of binary collisions(eqn.(10)) as a function of $N_{Coll}$ for
different $m_T-m_0$ intervals. The open squares represent the
experimentally extracted values, while the fully filled ones
indicate the GCM-based results.}
\end{figure}

Based on the analyses made by the chosen approach here, we make
the following particular observations:
\begin{enumerate}
\item[(i)] The modestly successful reproduction of the measured data
on $PP$ reaction, $PbPb$ and $PbNb$ collisions in the minimum bias events
confirm that the used basic models have good degree of reliability.\\
Besides, the efficacy of the model is exhibited in a somewhat
tangible way for the various degree of centrality of
nucleus-nucleus collisions demonstrated by Fig.~\ref{allclasses}.
In fact, the Fig.~\ref{modelratio} is a summary statement in
favour of the chosen model. \item[(ii)] The average $p_T$
behaviour is also explained by the present approach. But for any
specific nucleus-nucleus collision, the $s$-dependences of the
$<p_T>$ values of the various secondaries are to be obtained
mainly from the analysis of nucleon-nucleon collision in a
model-based manner. \item[(iii)] The ratios of the collisions with
various degrees of centrality of collisions to the minimum bias
events have been obtained with a fair degree of consistency.
\item[(iv)] The ratios of the most central to semi-central or
least central(peripheral) nuclear collisions have also been
computed by the used approach in conformity with the extracted
available data\cite{Aggarwal1} on pion production alone.
\end{enumerate}
\par
But, the above points cover no comments on the comparison of the
performances of the few chosen models which are under
consideration here and offer no insights into the state of
physical reality. Regarding the former, the Fig.7 illustrates
somewhat convincingly that the GCM performs much better to
accommodate data than both the PQCD and HIJING versions of the
standard multiparticle production models separately do. And,
concerning the latter, firstly, the status of anomalous nuclear
enhancement(ANE) is still ambiguous. In so far as pion production
in high energy $PbPb$ collision is concerned, this effect is seen
(Fig.7b) to be perceptible for the relatively peripheral
collisions. And for these non-central collisions the GCM describes
the observed effect modestly well. But the scenarios are different
for the other sets of the medium central to most central
collisions(Figs.7c and 7d), wherein the enhancement effect ceases
to exist and the reversals of nature occur quite prominently with
observations of the gradual and puzzling diminution of the effects
related with transverse mass. Thus, in any of such cases, the
concept of scaling with system size is not corroborated by the
extracted data obtained by WA98 collaboration\cite{Aggarwal1}.
True, all these data-trends are captured by the GCM somewhat
automatically. But the reasons for this spectacular success in
terms of collision dynamics, i.e. the number of collisions or the
number of participants, are still not very clear to us. Secondly,
yields of neutral pions per binary collisions could increase with
the number of collisions only upto a certain degree of
`hardness'($p_T \geq$ 1 GeV/c - 2 GeV/c) of the multiple
collisions. For large $p_T$-ranges, the validity of the physics of
parton saturation at larger transverse momentum region is quite
manifest at values of $N_{coll} \approx 200$(Figs.8c and 8d) from
both the extracted data-sets and GCM-based results.

\bigskip

\begin{center}
{\bf Acknowledgements}
\end{center}
\singlespacing{The authors are grateful to the concerned editor
and also to the learned referee for making some valuable comments
on an earlier version of the manuscript.}

\newpage
\begin{table}
    \caption{Different parameter values for $\pi^0$ production in $P+P$
    collision at 200 GeV}
    \begin{center}
            \begin{tabular}{lll}
         \hline
          $C_1$ & $p_0$(GeV/c) & $n$\\
         \hline
          $270\pm 5$ & $3.0\pm 0.2$ & $23\pm 1$\\
         \hline
        \end{tabular}
    \end{center}
    \label{tab: pp}
\end{table}
\begin{table}
    \caption{Necessary parameter values for neutral pion production in
    the minimum bias $Pb+Nb$ and $Pb+Pb$ collisions at 160A GeV.}
    \begin{center}
    \begin{tabular}{llll}
         \hline
          Collision & $C_3$ & $\alpha$(GeV/c)$^{-1}$ & $\beta$(GeV/c)$^{-2}$\\
         \hline
          $Pb+Nb$ & $(3.0\pm 0.5)\times 10^5$ & $0.17\pm 0.02$ & $0.033\pm 0.002$\\
      $Pb+Pb$ & $(8.6\pm 0.8)\times 10^5$ & $0.17\pm 0.02$ & $0.032\pm 0.001$\\
         \hline
        \end{tabular}
    \end{center}
    \label{tab: ABmb}
\end{table}
\begin{table}
 \caption{Various parameter values for $\pi^0$-production in $Pb+Pb$ collisions at $160A$ GeV for different centrality-values.}
    \begin{center}
        \begin{tabular}{llll}
         \hline
          $N_{Coll}$ & $C_3$ & $\alpha$(GeV/c)$^{-1}$ & $\beta$(GeV/c)$^{-2}$\\
         \hline
          10 & $19\pm 1$ & $0.13\pm 0.02$ & $0.042\pm 0.002$\\
      30 & $33\pm 2$ & $0.15\pm 0.02$ & $0.035\pm 0.003$\\
      78 & $130\pm 4$ & $0.14\pm 0.02$ & $0.038\pm 0.003$\\
          207 & $200\pm 6$ & $0.17\pm 0.01$ & $0.036\pm 0.003$\\
      408 & $250\pm 3$ & $0.16\pm 0.02$ & $0.030\pm 0.003$\\
          570 & $495\pm 5$ & $0.16\pm 0.01$ & $0.032\pm 0.004$\\
      712 & $746\pm 3$ & $0.14\pm 0.02$ & $0.035\pm 0.003$\\
      807 & $705\pm 8$ & $0.15\pm 0.01$ & $0.040\pm 0.003$\\
         \hline
        \end{tabular}
    \end{center}
    \label{tab: allclasses}
\end{table}
\begin{table}
  \caption{Used values of $<N_{part}>$ and $<N_{coll}>$ for various
centrality-classes\cite{Aggarwal1} of $PbPb$ collisions. The
values in column `Class' indicate vertically downwards gradual
transitions of the collisions from the-lowest-to-the-highest
centrality.}
   \begin{center}
   \begin{tabular}{llll}
   \hline
   Class & $E_T$(GeV) & $<N_{part}>$ & $<N_{coll}>$ \\
   \hline
   1 & $\leq 24.35$ & $12\pm2$ & $9.9\pm2.5$\\
   2 & $24.35-55.45$ & $30\pm2$ & $30 \pm 5$\\
   3 & $55.45-114.85$ & $63\pm2$ & $78 \pm 12$\\
   4 & $114.85-237.35$ & $132\pm3$ & $207\pm21$\\
   5 & $237.35-326.05$ & $224\pm1$ & $408\pm 41$\\
   6 & $326.05-380.35$ & $290\pm2$ & $569\pm 57$\\
   7 & $380.35-443.20$ & $346\pm1$ & $712\pm 71$\\
   8 & $>443.20$ & $380\pm1$ & $807\pm81$\\
   \hline
   \end{tabular}
  \end{center}
\end{table}

\newpage

\newpage
\begin{thebibliography}{*}
\bibitem{Aggarwal1} M.M.Aggarwal et al(WA98 Collaboration): Eur. Phys. Jour. {\bf C 23} (2002) 225.
\bibitem{De1} Bhaskar De, S. Bhattacharyya and P. Guptaroy: Jour. Phys. {\bf{G 28}}(2002)2963.
\bibitem{Faessler1} M.A.Faessler: Phys. Rep. {\bf 115}(1984) 1.
\bibitem{Peitzmann1} T.Peitzmann: Phys. Lett. {\bf B 450}(1999) 7.
\bibitem{Schmidt1} H.R.Schmidt and J.Schukraft: J. Phys. {\bf G 19}(1993) 1705.
\bibitem{Hagedorn1} R.Hagedorn : Riv. Nuovo. Cim. {\bf 6}(1983) 1.
\bibitem{Albrecht1} R.Albrecht et al(WA80 Collaboration): Eur. Phys. Jour. {\bf C 5}(1998) 255.
\bibitem{Antreasyan1} D.Antreasyan et al: Phys. Rev. {\bf D 19}(1979) 764.
\bibitem{SB1} S.Bhattacharyya: Lett. Nuv. Cim. {\bf 44}(1985)119.
\bibitem{De2} Bhaskar De, S.Bhattacharyya and P.Guptaroy: Accepted for publication in Int. Jour. Mod. Phys. {\bf A}(2002).
\bibitem{Hwa1} R.C.Hwa et. al.: Phys. Rev. {\bf C 64}(2001)054611.
\bibitem{Hove1} L.Van-Hove: Z. Phys. {\bf C 21}(1983)93.
\bibitem{Li1} S.Y.Li and X.N.Wang: Phys. Lett. {\bf B 527}(2002)85.
\bibitem{De3} Bhaskar De, S.Bhattacharyya and P.Guptaroy: Eur. Phys. Jour. {\bf A 16}(2003)415.
\bibitem{Adamus1} M.Adamus et al(EHS-NA22 Collaboration): Preprint, IFVE-86-210 (1986).
\bibitem{Donaldson1} G.Donaldson et al: Phys. Rev. Lett. {\bf 36}(1976) 1110.
\end{thebibliography}
\end{document}